# Cross Sections of Coronal Loop Flux Tubes

James A. Klimchuk

NASA Goddard Space Flight Center

Craig E. DeForest

Southwest Research Institute

## Abstract

Coronal loops reveal crucial information about the nature of both coronal magnetic fields and coronal heating. The shape of the corresponding flux tube cross section and how it varies with position are especially important properties. They are a direct indication of the expansion of the field and of the cross-field spatial distribution of the heating. We have studied 20 loops using high spatial resolution observations from the first flight of the Hi-C rocket experiment, measuring the intensity and width as a function of position along the loop axis. We find that intensity and width tend to either be uncorrelated or to have a direct dependence, such that they increase or decrease together. This implies that the flux tube cross sections are approximately circular under the assumptions that the tubes have non-negligible twist and that the plasma emissivity is approximately uniform along the magnetic field. The shape need not be a perfect circle and the emissivity need not be uniform within the cross section, but sub-resolution patches of emission must be distributed quasi-uniformly within an envelope that has an aspect ratio of order unity. This raises questions about the suggestion that flux tubes expand with height, but primarily in the line-of-sight direction so that the corresponding (relatively noticeable) loops appear to have roughly uniform width, a long-standing puzzle. It also casts doubt on the idea that most loops correspond to simple warped sheets, although we leave open the possibility of more complex manifold structures.

## 1. Introduction

It has been recognized for many years that coronal loops---thin curved intensity features in coronal images---have widths that are approximately uniform along their length (Klimchuk et al. 1992). This has been a major puzzle, since the loops are thought to coincide with magnetic flux tubes, and the magnetic field must expand with height in the corona, at least on average. Why would these particular flux tubes not expand, and why would they have enhanced brightness? The low-$\beta$ corona is completely filled with field. Individual flux tubes are visible only because they have different temperature and/or density from their surroundings. Since the thermodynamic properties of the plasma are directly related to the energy deposited in the tube, the answers to these questions hold important clues about the nature of coronal heating itself.

The term "loop" is often used for both observational features and flux tubes. To avoid confusion, we will reserve "loop" for the two-dimensional structures in images and use "flux tube" for the corresponding structures in three-dimensional space. A loop is the projection of a flux tube onto the image plan, or plane of the sky. Any expansion of the flux tube along the line of sight does not affect the width of the loop.

Klimchuk, Antiochos, & Norton (2000) offered one possible explanation for the mystery of uniform widths. They suggested that loops correspond to locally twisted flux tubes. The well-known pinch effect would cause the cores of such tubes to constrict. However, since line tying prevents constriction in the photosphere, the effect should increase with altitude, thereby reducing the expansion of the field compare to the corresponding untwisted tube. Since twist is associated with electric current, enhanced coronal heating might also be expected, causing the tube to be brighter than its surroundings. Force-free magnetic field models were constructed to confirm that differential constriction in a twisted tube does indeed promote width uniformity. However, so much twist would be required to match the observations that the tubes would be kink unstable, and the idea was rejected.

Other more recent ideas for explaining the uniform widths appeal to the shape of the tube cross section. It is often assumed, without justification, that the cross section is circular. Though not ruled out, there is no obvious reason why the heating should itself be axially symmetric. Symmetry could, in principle, arise from the spreading of the energy along the field. If the field lines have wandering trajectories, as might be expect from the chaotic photospheric driving, then they might fill an axially symmetric envelope (Klimchuk 2000, Fig. 22b). However, it is challenging to understand how the axial symmetry would be maintained along the full length of the tube. In general, the shape of the cross section changes along a tube, so that a circle at one position would transition into, e.g., an oval at another position. Examples from force-free models of active regions are given in Lopez Fuentes, Klimchuk, & Demoulin (2006). It is worth noting that localized twist would tend to keep the cross section circular, due the tension force in the azimuthal component of the field (Klimchuk et al. 2000).

Malanushenko & Schrijver (2013) suggested that non-circular cross sections might reconcile the apparent inconsistency between the observed uniform loop widths and the expected expansion of the field. There is no contradiction if the expansion occurs preferentially along the line of sight (LOS), rather than within the image plane (IP) where it is measured. A natural question that arises is why flux tubes should expand preferentially along the LOS. Surely the location of the observer has no bearing on the structure of the corona. Malanushenko & Schrijver proposed that a selection bias is at work. Because the corona is optically thin, loops expanding along the LOS will be brighter and more noticeable than loops expanding within the IP. More specifically, the brightness will decrease with height less quickly in the former, since any gravitational stratification will be offset by the increasing line-of-sight depth. Warm (~1 MK) loops are indeed observed to have scale heights greater than expected for hydrostatic equilibrium at the nominal temperature (Aschwanden, Schrijver, & Alexander 2001). Steady flows do not change this conclusion (Patsourakos, Klimchuk & MacNeice 2004). Line-of-sight expansion is one explanation for the super-hydrostatic scale heights, but there are others (e.g., DeForest 2007).

The Malanushenko & Schrijver idea that the most noticeable loops expand preferentially along the LOS is very appealing, but it needs to be tested. One test would be to determine whether expanding loops that fade with height exist in the expected numbers given the quantity of non-expanding loops that fade substantially less. The properties of the former group can be predicted from the observed properties of the latter by assuming an expansion along the LOS that is consistent with expectations from magnetic field models.

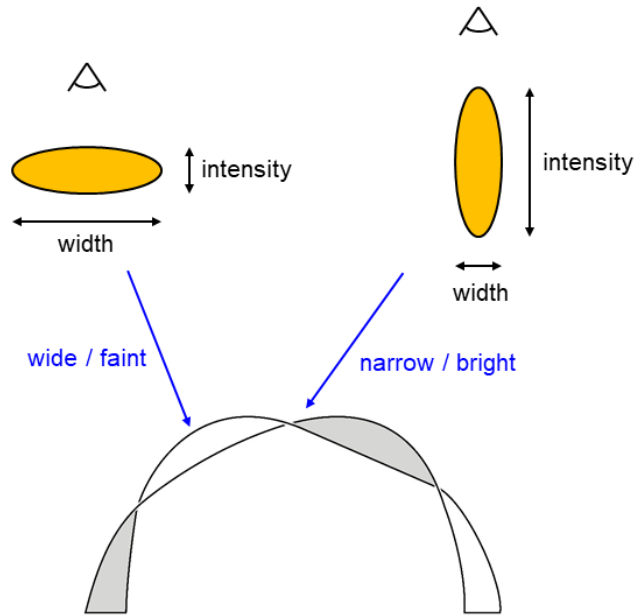

**Figure 1.** Idealized sketch of a coronal loop corresponding to a twisted flux tube with a non-circular cross section showing how the intensity and width and are anti-correlated. Bottom part adopted from Klimchuk (2000).

We report here on a different test of the Malanushenko & Schrijver idea. Coronal flux tubes are expected to be twisted, as discussed in Section 5. Each tube can be a self-contained twisted structure or simply a subset of field lines within a larger structure that is twisted on a scale greater than the tube diameter. In either case, the cross section of the tube rotates as a function of position along the tube axis. This has important observational consequences. For any non-circular cross section, such as an oval, there will be places where the LOS is aligned with the long dimension, and other places where it is aligned with the short dimension. The loop will appear relatively bright and narrow at the first case, and relatively faint and wide in the second. Thus, intensity and width are expected to be anti-correlated. This is shown schematically in Figure 1. Klimchuk (2000) pointed out that the approximate uniformity of width in observed coronal loops suggests that the cross sections are roughly circular. The study we report here investigates the predicted anti-correlation between intensity and width.

## 2. Observations

Since we want the best possible measurements of loop width, we use high spatial resolution observations from the first flight of the Hi-C sounding rocket, which took place on 2012 July 11. Normal incidence images were obtained with 2 s exposures in a narrow wavelength band centered at 193 Å. The passband is similar to that of the 193 Å channel of the Atmospheric Imaging Assembly (AIA) on the Solar Dynamics Observatory (SDO) and is dominated by a line of Fe XII, which is formed at about 1.5 MK under equilibrium ionization conditions. The 6.8x6.8 arcmin$^2$ field of view captured Active Region 11520, located [-150, -281] arcsec from disc center. The spatial resolution of the observations is estimated to be 0.3-0.4 arcsec, which includes three contributions: a finite pixel size of 0.10 arcsec, a point spread function with a full width at half maximum of 0.09 arcsec, and spacecraft jitter. The spatial resolution of AIA/SDO is 3-5 times coarser. Further details on the instrument and observations can be found in Kobayashi et al. (2014) and Winebarger et al. (2014).

Figure 2 shows Hi-C frame number 23 (solar image sequence), taken at 18:54:17.79 UT that was used for our study. The image can also be found at https://hic.msfc.nasa.gov/gallery.html. One of us (C.D.) identified 20 loops, indicated in the figure, and produced "riverine" plots, which are extracted sub-images in which the loop has been approximately straightened. The other of us (J.K.) analyzed those plots using the method described in Klimchuk et al. (1992) and Klimchuk (2000) and summarized below. The loops were selected using visual recognition of distinguishable features, with an attempt to encompass a broad variety of morphologies and lengths. Because the selection is "by eye" and not algorithmic, the results should not be used for quantitative statistics---but the ensemble is suitable to identify trends. The loops were visible for the full 200 s duration of solar observations, which is not surprising given that loops at these temperatures have typical lifetimes of 1000-5000 s (Klimchuk, Karpen, & Antiochos 2010).

To obtain the most reliable measurements, we only consider those sections of the loops that are free of especially bright and complex background emission. The usable section usually constitutes a majority of the total loop length, but not always a large majority. In some cases, it is not possible to identify where in the photosphere the flux tube terminates, so it is not possible to know what fraction of the total tube length the section represents. All but two of the loops have a monolithic appearance. While all loops may be multi-stranded on a scale below the Hi-C resolution, Loops 16 and 19 exhibit clearly detectable sub-structure and are excluded from the results.

Two of the remaining eighteen loops are part of an earlier study of four loops that compared the widths measured independently with Hi-C and AIA/SDO using contemporaneous observations (Klimchuk 2015, Section 9). The widths are similar in the two images, indicating that the loops, or more specifically their outer envelopes, are resolved by AIA and well resolved by Hi-C. A similar conclusion was reached by Peter et al. (2013). As noted, we believe that substructure may exist, but well below the resolution of Hi-C. All four of the previously measured loops are included in our results and are designated JK1-4. Loops 9 and JK4 are the same, as are loops 14 and JK3, but the measured sections from the two studies have differing length and only partly overlap.

A recent paper reports 172 Å (Fe IX, T ~ 0.8 MK) observations of an active region made during the third flight of Hi-C (Williams et al. 2020). That active region has many more identifiable loops than the one studied here, which is rather unusual in its dearth of loops. A 60 s time-averaged image was compared with a corresponding 171 Å image from AIA, and a number of loops were selected for detailed study. Many of the loops appear similar in the two images and have similar cross-axis intensity profiles, while a sizable number show clear evidence of substructure.

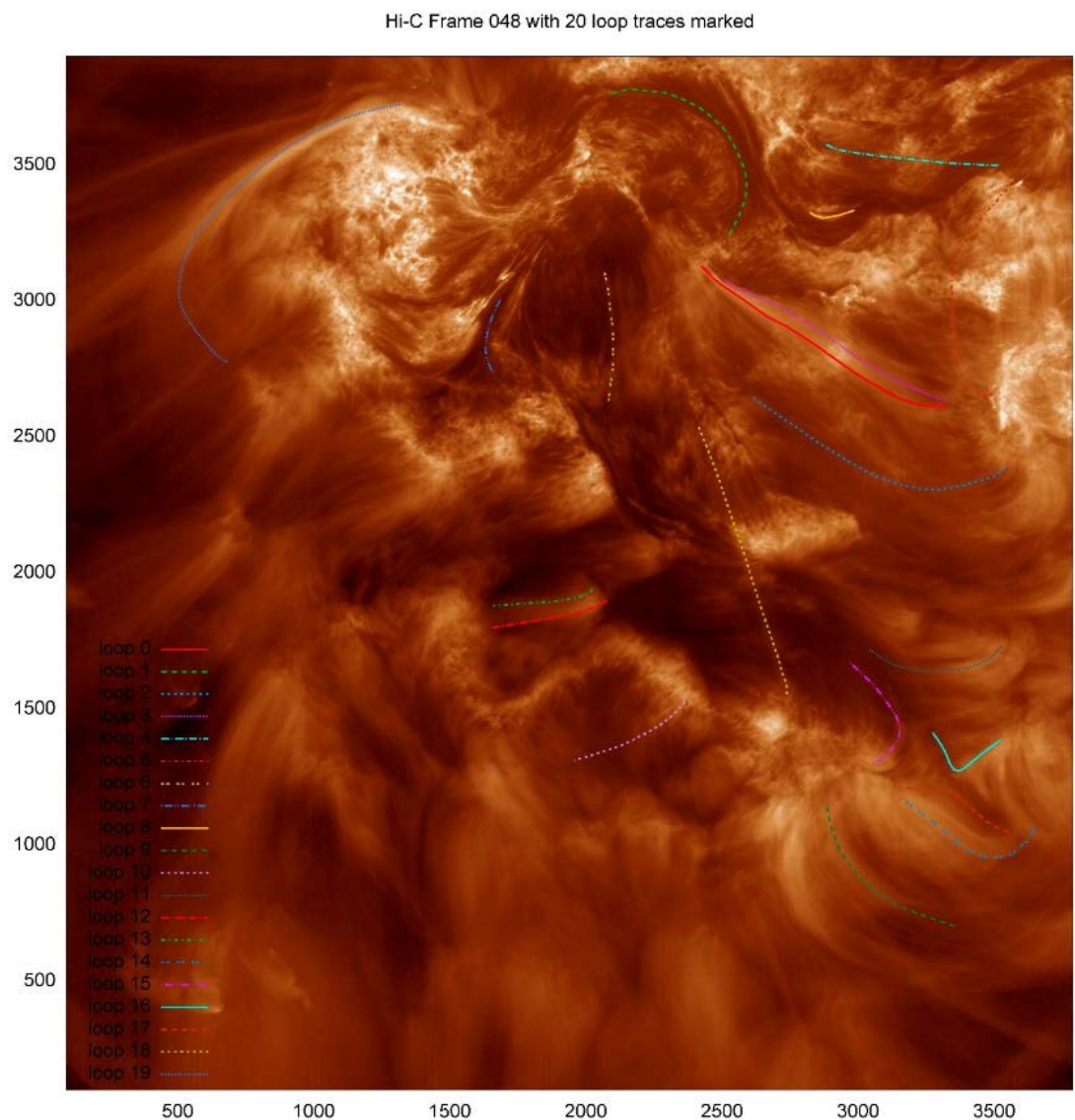

**Figure 2.**  Hi-C rocket image indicating the loops used in this study. Measurements of Loop 14 (lower right, blue double dot, extending closest to the right edge) are shown in Figures 3-5. The image is 290 Mm on a side.

## 3. Analysis

The first step of the analysis is to further straighten the loops in the riverine plots (Loops JK1-4 were taken directly from the original image and not from riverine plots). This is accomplished by selecting points along the loop axis and fitting them with a polynomial, usually third order. This defines a new coordinate system onto which the original image is mapped. An example is shown in the leftmost panel of Figure 3. This is Loop 14 from the lower-right region in Figure 2 (double dotted blue curve). As mentioned, this is also Loop JK3 from Klimchuk (2015) and one of the loops examined in Peter et al. (2013, Figs. 3 and 6).

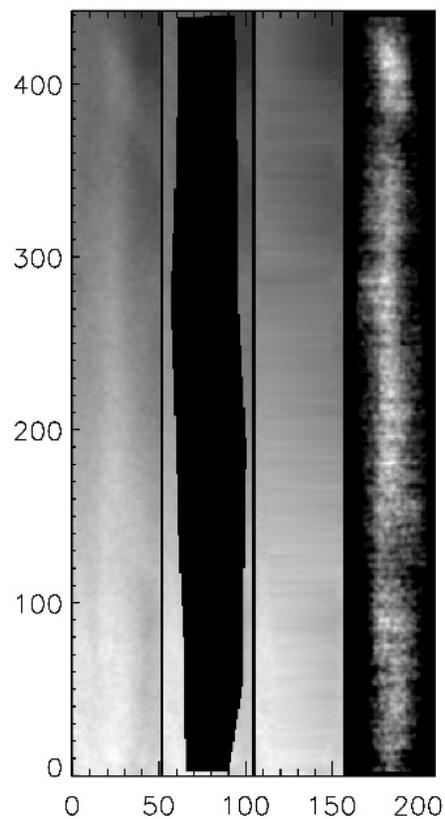

**Figure 3.** Loop 14, left to right: straightened loop image; with loop subtracted; linearly interpolated background; background-subtracted loop. Spatial units are pixels, where one pixel corresponds to 75 km.

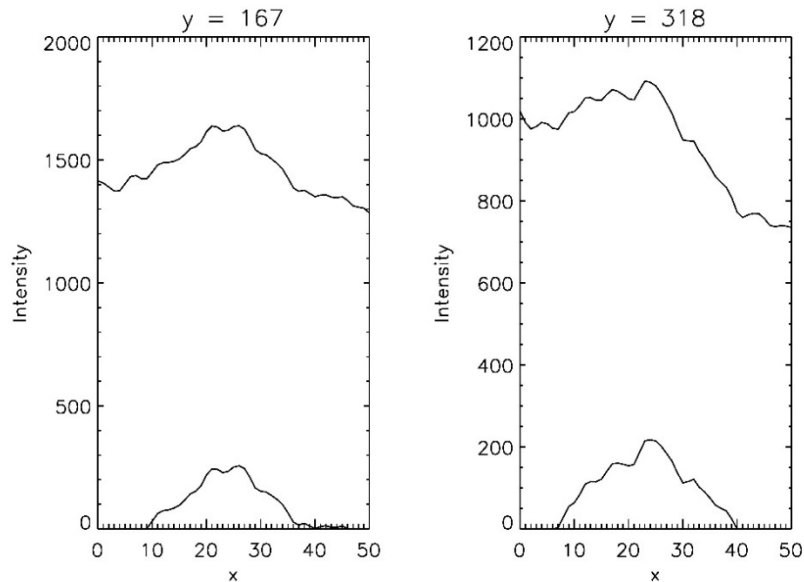

**Figure 4.** Cross axis intensity profiles from Loop 14, with and without background, at axis positions 167 (left) and 318 (right) in Figure 3.

The next step is to subtract the background emission, which is usually considerably brighter than the loop itself. The loop is subjectively identified and removed (second panel in Fig. 3), and intensity is linearly interpolated between the left and right edges of the gap (third panel). The background image is then subtracted from the original straightened image, leaving the loop itself (fourth panel).

Figure 4 shows cross-axis intensity profiles before and after background subtraction at two locations along the loop axis. These are horizontal rows extracted from the first and fourth panels in Figure 3 at $y = 167$ and 318. We selected these locations because they are representative of relatively uniform (flat) and nonuniform (slanted) backgrounds. The loops in our dataset are typically only 10-50% as bright as the background, so background subtraction is critical. To be conservative, we have excluded from our results any locations where the peak of the intensity profile does not exceed 25 normalized Hi-C detector counts[1] after background subtraction. These are relatively uncommon, and in no case do they account for more than 7% of the analyzed section.

We measure the loop width by computing the second moment of background-subtracted intensity profile, i.e., its standard deviation. This has the advantage over other methods in that it makes no assumption about the shape of the profile. Some authors use Gaussian fits, but there is no compelling reason for doing so. In fact, a circular cross section of uniform emissivity would produce a distinctly non-Gaussian profile, just as we find in Figure 4. For a circular cross section, the diameter is four times the standard deviation, so a standard deviation of 6 pixels would correspond to a diameter of 1800 km. We do not attempt to correct for the finite spatial resolution of the observations, as has been done with Transition Region and Coronal Explorer (TRACE) observations (Watko & Klimchuk 2000; Lopez et al. 2006) and Yohkoh Soft X-ray Telescope (SXT) observations (Klimchuk 2000). There is no significant need to do so, since, in contrast to the earlier observations, the loops are many times wider than the resolution.

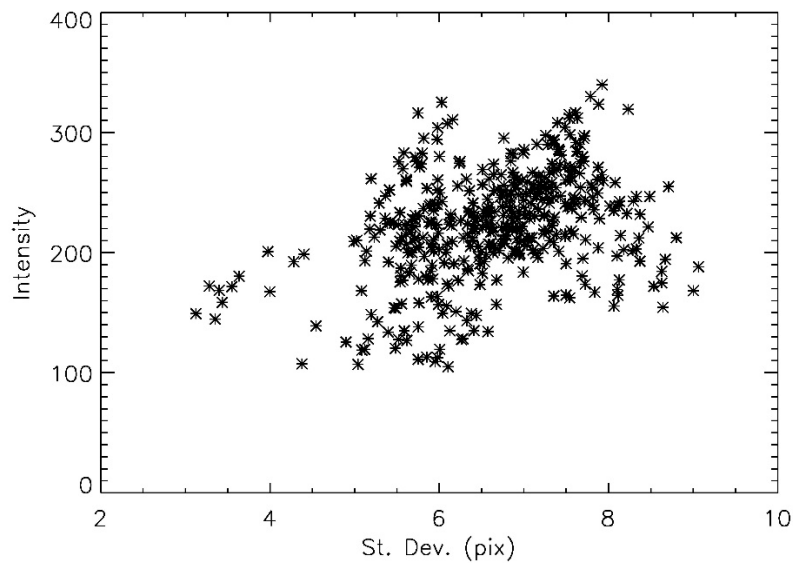

**Figure 5.** Peak intensity versus standard deviation (width) from the intensity profiles along Loop 14.

---

[1] Detector counts were normalized using a flat field function derived post facto from the entire Hi-C image set including darks, with a modal value of 1: they therefore represent original detector "digitizer number" counts, corrected for known dark current, vignetting, and flat-field variations. These counts are proportional, but not identical, to the number of photons detected during the exposure.

Figure 5 is a scatter plot of peak intensity versus standard deviation (width) at 438 positions along Loop 14. Visually, there appears to be a positive correlation, in contradiction to the anti-correlation predicted for non-circular cross sections. To verify this impression, we perform a rigorous statistical analysis consisting of two parts. First, we determine the strength of the correlation, and second, we determine the sense and strength of the dependence between the two variables. We assume a power law relationship of the form $I \propto w^{\alpha}$, where $I$ is the peak intensity and $w$ is the width. A strong correlation means that the $(w, I)$ data points are tightly clustered along a straight line in a log-log plot. The sense and strength of the dependence are given by the sign and magnitude of the line's slope, i.e., the sign and magnitude of $\alpha$. Positive $\alpha$ indicates a direct correlation and negative $\alpha$ indicates an inverse, or anti, correlation.

We adopt a nonparametric statistical approach. Specifically, we use a rank ordering method based on the weighted t-statistic (Efron & Petrosian 1992), as described in Porter & Klimchuk (1995). The advantage of this method is that it makes no assumption about the underlying distribution of the measurements. Standard techniques, such as the familiar correlation coefficient, assume that the meaurements are normally distributed about the true value, i.e., that errors in the measurements are represented by a Gaussian distribution. It is unknown whether that applies to our width and intensity measurements.

The weighted t-statistic is a carefully designed number computed from the $(w, I)$ data pairs. Larger values indicate a stronger correlation. The value computed from Figure 5 is such that the probability of obtaining that value or higher from random data is only $8.7 \times 10^{-6}$. Thus, we can be very confident that the data are, in fact, correlated.

The most probable value of $\alpha$ is the one that maximizes the likelihood of correlation between $I$ and $w^\alpha$, which we also determine using the weighted t-statistic. The most probable $\alpha$ for Loop 14 is 0.34, indicating a direct dependence (positive correlation) of modest strength. The 90% confidence interval is [0.24, 0.44], meaning that there is a 90% probability that $\alpha$ falls within this range. Further details of the method can be found in Porter & Klimchuk (1995).

## 4. Results

Table 1 gives results for the 18 "monolithic" loops measured in this study and 4 loops measured previously. The first loop is Loop 0. Loops JK1-4 are also named Loops 20-23. As mentioned, Loops 16 and 19 are excluded because they have clearly visible substructure. The columns in the table list, from left to right, the loop number, probability that $I$ and $w$ are random (a small value indicates a high likelihood of correlation), most probable $\alpha$, 90% confidence interval for $\alpha$, average of the background-subtracted peak intensity in normalized instrument counts, average of the loop-to-background intensity ratio, and length of the measured loop section in pixels.

Figure 6 is plot of the most probable $\alpha$, henceforth simply referred to as $\alpha$, versus loop number. The error bars give the 90% confidence intervals. Blue indicates < 5% probability that the data are random, while red indicates > 5% probability. Thus, blue loops have a high likelihood that intensity and width are correlated, and red loops are consistent with no correlation. Because the determination of $\alpha$ is meaningful only when there is a substantial likelihood of correlation, the $\alpha$ values for the red loops must be treated with caution. It is perhaps reassuring that they are consistent with 0 to within the errors. We can say with confidence that red loops have a very weak dependence between intensity and width, if one exists at all. The squares and triangles differentiate measured loop sections that are longer and shorter than 250 pixels, respectively.

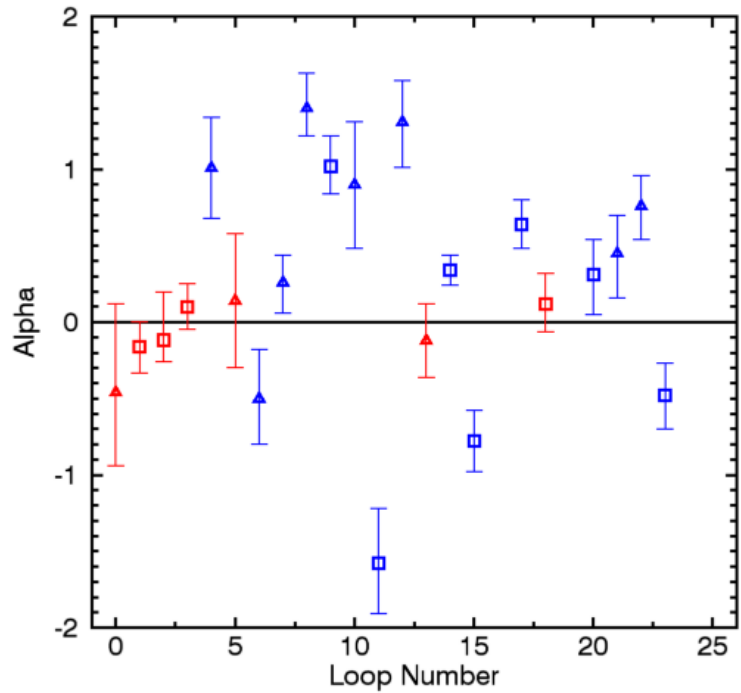

*Figure 6.* Most probable $\alpha$ for the loops in Table 1. Error bars indicate 90% confidence intervals. Blue indicates a high likelihood of correlation; red is consistent with no correlation. Squares and triangles indicate measured loop sections longer and shorter than 250 pixels, respectively.

What stands out most in Figure 6 is the dominance of loops with a positive intensity-width correlation. Eleven of the blue loops have $\alpha > 0$, compared to four with $\alpha < 0$. That represents a nearly three-to-one imbalance. Only one loop has $\alpha < -0.8$, compared to five with $\alpha > 0.8$.

To assess whether background subtraction may systematically compromise the loop width measurements, we examined the correlation between width and background intensity. We computed a power-law index for the dependence of background intensity on width, directly analogous to the $\alpha$ for loop intensity. Figure 7 plots the loop $\alpha$ on the y-axis against the background $\alpha$ on the x-axis. As in Figure 6, red and blue indicate relatively low and high probability of correlation with loop intensity. Squares and triangles now indicate relatively low and high probability of correlation with background intensity.

There are two things to note. First, the magnitudes of $\alpha$ are much smaller for background intensity than for loop intensity, suggesting that any dependence of width on background intensity is weaker than the dependence on loop intensity. Second, a majority the background $\alpha$ are negative, perhaps indicating that systematic errors associated with background subtraction tend to diminish the positive correlation between loop intensity and width.

Lopez Fuentes, Demoulin, & Klimchuk (2008) examined the effect of background subtraction on loop properties measured with TRACE. They concluded that brighter backgrounds can cause an underestimate of both width and intensity. This would increase $\alpha$ for the loop compared to its true value. We suggest that the effect is weaker for Hi-C observations, since the loops are several times wider than the spatial resolution, in contrast to TRACE. Also, we find no clear trend in the correlation between loop intensity and background intensity in the Hi-C data. Note that the widths presented in Lopez Fuentes et al. are four times the standard deviation of the intensity profile, which is the measure we use here.

To evaluate whether the subjective identification of the loop region used to determine the background has a significant effect on the results, we repeated the measurements with a gap (e.g., second panel in Fig. 3) that is 3 pixels wider on each side. The new $\alpha$ values are shown in Figure 8. As can be seen, the basic results are unchanged.

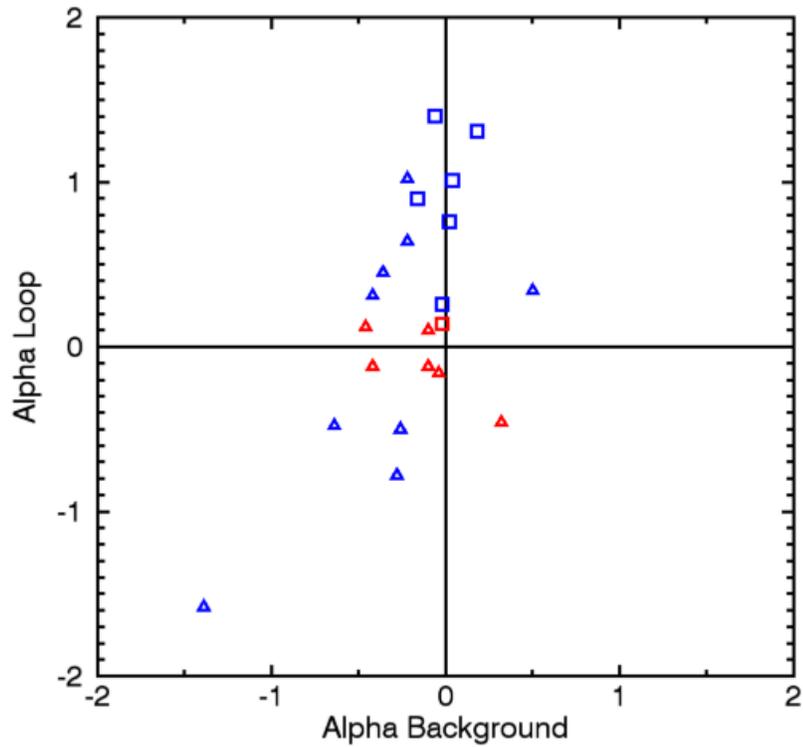

**Figure 7.** *Most probable $\alpha$ for the loop versus the corresponding $\alpha$ for the dependence of width on background intensity. Blue indicates a high likelihood of correlation with loop intensity; red is consistent with no correlation. Triangles indicate a high likelihood of correlation with background intensity; squares are consistent with no correlation.*

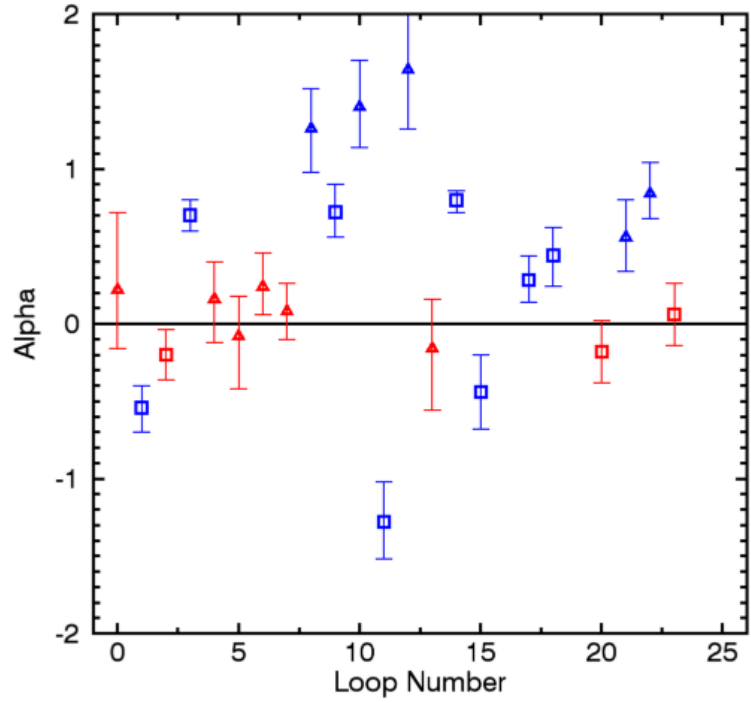

*Figure 8*.  *Same as Figure 6, but with a loop region (gap) used to determine the background emission that is 3 pixels wider on each side.*

| Loop Number | Uncorrelated Probability | Most Probable $\alpha$ | 90% Confidence Interval | Loop Intensity (counts) | Loop/Bkg Intensity Ratio | Length (pix)$^2$ |
|---|---|---|---|---|---|---|
| 0 | 0.19 | -0.46 | [-0.94, 0.12] | 669 | 0.43 | 181 |
| 1 | 0.10 | -0.16 | [-0.33, 0.00] | 89 | 0.19 | 821 |
| 2 | 0.20 | -0.12 | [-0.26, 0.20] | 68 | 0.13 | 506 |
| 3 | 0.27 | 0.10 | [-0.05. 0.25] | 198 | 0.16 | 828 |
| 4 | $< 10^{-5}$ | 1.01 | [0.68, 1.34] | 175 | 0.23 | 171 |
| 5 | 0.61 | 0.14 | [-0.30, 0.58] | 180 | 0.17 | 104 |
| 6 | $9.7 \times 10^{-3}$ | -0.50 | [-0.80, -.18] | 94 | 0.39 | 177 |
| 7 | 0.043 | 0.26 | [0.06, 0.44] | 61 | 0.20 | 218 |
| 8 | $< 10^{-5}$ | 1.40 | [1.22, 1.63] | 816 | 1.19 | 138 |
| 9 | $< 10^{-5}$ | 1.02 | [0.84, 1.22] | 180 | 0.17 | 253 |
| 10 | $1.70 \times 10^{-4}$ | 0.90 | [0.48, 1.31] | 231 | 0.26 | 106 |
| 11 | $< 10^{-5}$ | -1.58 | [-1.91, -1.22] | 284 | 0.32 | 306 |
| 12 | $< 10^{-5}$ | 1.31 | [1.01, 1.58] | 226 | 0.43 | 146 |
| 13 | 0.47 | -0.12 | [-0.36, 0.12] | 130 | 0.23 | 82 |
| 14 | $< 10^{-5}$ | 0.34 | [0.24, 0.44] | 223 | 0.21 | 438 |
| 15 | $< 10^{-5}$ | -0.78 | [-0.98, -0.58] | 74 | 0.23 | 340 |
| 17 | $< 10^{-5}$ | 0.64 | [0.48, 0.80] | 97 | 0.14 | 259 |
| 18 | 0.27 | 0.12 | [-0.06, 0.32] | 106 | 0.40 | 321 |
| JK1 (20) | $4.9 \times 10^{-2}$ | 0.31 | [0.05, 0.54] | 162 | 0.48 | 804 |
| JK2 (21) | $9.3 \times 10^{-3}$ | 0.45 | [0.16, 0.70] | 311 | 0.28 | 168 |

---

$^2$ 100 pixels correspond to 7.5 Mm.

| | | | | | | |
|---|---|---|---|---|---|---|
| JK3 (22) | $< 10^{-5}$ | 0.76 | [0.54, 0.96] | 219 | 0.21 | 190 |
| JK4 (23) | $2.4 \times 10^{-4}$ | -0.48 | [-0.70, -0.27] | 199 | 0.22 | 293 |

## 5. Discussion

It is clear from Table 1 and Figures 6, 7, and 8 that the measured loops do not follow the trend expected for flux tubes with significantly non-circular cross sections and finite twist. For simple cross-sectional shapes, such as the oval in Figure 1, intensity should vary inversely with width to the first power, i.e., $\alpha = -1$. Instead, we find that a sizable majority of the loops either have a clear positive correlation, $\alpha > 0$, or are consistent with no correlation. We are forced to conclude that, for these loops, either the cross section is approximately circular (aspect ratio near unity), the tubes are untwisted, or the loops do not in fact correspond to flux tubes. We return to the third option below. Of the remaining two, it seems unlikely that a sizable majority of the tubes would be untwisted, leaving circular cross sections as the more likely explanation. The cross section need not be a true circle. It could be quite irregular as long as the dimension is comparable in all directions.

Two other studies came to the same conclusion of circular cross sections, but using different approaches that do not rely on twist. West, Klimchuk, & Zhukov (2014) examined eleven loops observed at quadrature by the Extreme Ultraviolet Imagers on the two STEREO spacecraft. The widths were measured to be comparable from the two orthogonal viewing angles, though the uncertainties are large.

Kucera et al. (2019) compared the widths of two loops with their line-of-sight thicknesses as inferred from emission measures and densities determined using spectra from the Extreme

Ultraviolet Imaging Spectrometer on Hinode. For an assumed filling factor near unity, the observations are consistent with a circular cross section. Uncertainties are large, however, and significantly non-circular shapes are also allowed by those data. Smaller filling factors imply greater aspect ratios, but large ratios are ruled out by contemporaneous observations from STEREO-A. Whatever its value, if the filling factor does not vary significantly with position along the loops, as we would not expect it to due to the low-$\beta$ nature of the corona, then the loops do not expand greatly with height in the LOS direction.

Our conclusion that the loops in our study have approximately circular cross sections rests on the assumption that they are twisted enough to influence the measured widths and intensities. Do we expect a substantial twist? As discussed in the introduction, the twist can be localized, giving rise to a classical twisted flux tube, or it can be a large-scale twist, in which case the flux tube is simply a subset of field lines within a much bigger magnetic structure. In either case, the cross section rotates with position along the tube. There is an abundance of direct and indirect evidence that the coronal magnetic field is twisted over a wide range of scales, as we now discuss.

Correlation tracking applied to high-resolution photospheric magnetograms of active regions (longitudinal field component) reveals a complex velocity pattern that would produce a high level of coronal twist on scales of one to several thousand kilometers and likely even smaller (Yeates, Hornig, & Welsch 2012). Maps of the measured vector magnetic field in the active region photosphere also imply small-scale coronal twist (Welsch 2015). Finally, magnetoconvection simulations indicate photospheric driving that, once again, would produce coronal twist on scales of one to several kilometers in active regions (Candelaresi et al. 2018; Shelyag et al. 2011).

Evidence for twist on larger scales is even more compelling. As reviewed by Toriumi & Wang (2019) and van Driel-Gesztelyi & Green (2015), this evidence includes sheared polarity inversion lines, soft X-ray sigmoids, twisted sunspots, extrapolations of photospheric vector magnetic fields, and comparisons of coronal and chromospheric features with linear force-free extrapolation models based on the measured longitudinal field.

While highly suggestive, the generic tendency for active region magnetic fields to be twisted on a variety of scales does not prove that the flux tubes in our study have sufficient twist to support the inference of circular cross sections. Detailed magnetic field models of this specific active region, number 11520, would be very helpful to strengthen the case. Thalman, Tiwari, & Wiegelmann (2014) used a nonlinear force-free model to study a well-known braided feature in the Hi-C observations (Cirtain et al. 2013), and something similar could be done for our loops.

The lack of a clear trend for intensity and width to be inversely correlated with $\alpha$ = -1 suggests circular cross sections, but how can we understand the observed positive correlations ($\alpha$ > 0)? A positive correlation would be produced by a circular cross-section if its area varied along the tube, assuming approximately uniform plasma emissivity along the field. A non-circular cross section of varying area would also produce a positive correlation if the shape remained constant *and* there were no significant twist. Local variations in flux tube area are not unexpected, especially in the vicinity of topological features such as null points. Consistent with earlier findings, there is no clear evidence that our loops expand systematically with height, even though the coronal magnetic field must do so on average. A proper study must be performed, however.

In summary, we find that, for each of 20 loops from the first Hi-C rocket flight, intensity and width tend to either be uncorrelated or to have a direct dependence, such that they increase or decrease together along the loop. This implies that the cross sections of the corresponding flux tubes are approximately circular under the assumption that the tubes have non-negligible twist and approximately uniform emissivity along the field. It is not required that the cross section be a perfect circle. It could be highly irregular, but the aspect ratio cannot be greatly different form unity. It is also not required that the emissivity be uniform within the cross section. It could be concentrated it sub-resolution patches, which themselves could have large aspect ratios, but the distribution of patches must be approximately uniform. Figure 9 is an example of an emission pattern that fits our definition of a circular cross section. This picture is consistent with the idea that loops are comprised of multiple interacting strands. Explaining the collective behavior that gives rise to loops, including their circular cross sections, must be part of any successful coronal heating theory.

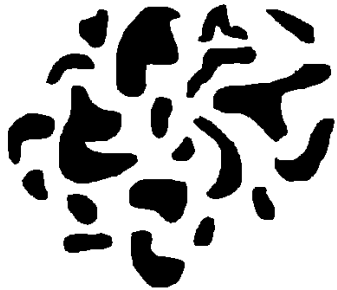

**Figure 9.** *Example of an emission pattern in a flux tube cross section that the fits our definition of circular.*

Our results raise questions about the appealing interpretation offered by Malanushenko & Schrijver (2013) for uniform loop widths. Paraphrased, that interpretation is that loops are associated with flux tubes that expand with height, as expected, but preferentially in the LOS direction, where it is not revealed as a changing loop width. However, if this interpretation were correct, the bright flux tubes would not have circular cross sections, except perhaps at a single location in one or both legs. We do not wish to suggest that our inference of circular cross sections is definitive or applies universally to all loops. It seems likely that the Malanushenko & Schrijver interpretation is correct for many loops, but what fraction of loops is an open question. The mystery of uniform widths has not yet been fully solved. We suggest that progress can be made on several fronts moving forward.

First, intensity-width correlations should be examined in data from the most recent Hi-C flight (Rachmeler et al. 2019). The active region observed during that flight has many more distinctive loops than were present during the first flight, and this should greatly improve the statistics.

Second, magnetic field extrapolation models should be constructed for the active regions from both flights to estimate the amount of twist in the flux tubes. If the twist is minimal, it would greatly affect our inference of circular cross sections.

Third, synthetic images from MHD simulations should be analyzed in a similar manner to actual data. These simulations have become impressively realistic in their appearance, including the presence of many loops (Rempel 2017). Do these loops obey the same intensity-width correlations as observed loops? Do they have a similar lack of expansion with height? In addressing these questions, it is critical that the models have a level of twist that is representative of real active regions. The present versions likely do not.

Malanushenko et al. (2020) recently studied an MHD simulation of an active region and found that some loops in the synthetic images correspond to "veils" of enhanced emissivity within the 3D volume. These can be thought of as flux tubes of extreme aspect ratio, though they are nothing like conventional, or tubular, flux tubes. Only those parts of the veil that are aligned with the LOS produce noticeable loops. Since veils will be warped in a field with twist, the loops that they produce should exhibit anti-correlated intensity and width, just as conventional flux tubes do. Is this the case in the synthetic images? Before drawing any conclusions, it must first be verified that the model field has realistic twist. Also, we note that some loops in the synthetic images from Malanushenko et al. (2020) come from localized volumes of strongly enhanced emissivity, rather than distributed veils of modestly enhanced emissivity with the right orientation, i.e., they correspond to flux tubes with approximately circular cross sections.

Finally, Peter & Bingert (2012) have suggested that cross-field temperature gradients result in loops that do not correspond precisely with flux tubes in observations having a narrow temperature sensitivity, including those from AIA and Hi-C. A loop might not appear to expand with height in an expanding field if the cross-field temperature gradient increases systematically toward the apex. It is difficult to imagine what would produce such a systematic thermodynamic structure throughout an active region, so we suggest that this can explain at most a subset of observed loops. The interpretation can be tested by looking for the expected spatial offset in loops observed in different temperature channels.

It seems that a universal explanation of coronal loops is still lacking. Perhaps multiple causes, at least one not yet identified, are at work. We must continue to investigate this fascinating phenomenon, as it holds vital clues about the nature of coronal magnetic fields and coronal heating.

We acknowledge fruitful discussions on the topic of loop structure with Therese Kucera, Anna Malanushenko, and Marcelo Lopez Fuentes, and we thank Therese and Anna for commenting on the manuscript. We also thank the referee for helpful comments. This work was supported by the NASA Heliophysics Guest Investigator program (NNX16AG98G) and the GSFC Internal Scientist Funding Model (competitive work package) program. We acknowledge the High resolution Coronal Imager instrument team for making the flight data publicly available. MSFC/NASA led the mission and partners include the Smithsonian Astrophysical Observatory in Cambridte, Mass.; Lockheed Martin's Solar Astrophysical Laboratory in Palo Alto, Calif.; the University of Lancashire in Lancashire, England; and the Lebedev Physical Institute of the Russian Academy of Sciences in Moscow.

## References

Aschwanden, M. J., Schrijver, C. J., & Alexander, D.  2001, ApJ, 550, 1036

Candelaresi, S., Pontin, D. I., Yeates, A. R., Bushby, P. J., & Hornig, G.  2018, ApJ, 864, 157

Cirtain, J. W., Golub, L., Winebarger, A. R., et al.  2013, Natur, 493, 501

DeForest, C. E.  2007, ApJ, 661, 532

Efron, B., & Petrosian, V.  1992, ApJ, 399, 345

Klimchuk, J. A.  2000, SoPh, 193, 53

Klimchuk, J. A.  2015, RSPTA, 373, 20140256

Klimchuk, J. A., Antiochos, S. K., & Norton, D.  2000, ApJ, 542, 504

Klimchuk, J. A., Karpen, J. T., $ Antiochos, S. K.  2010, ApJ, 714, 1239

Klimchuk, J. A., Lemen, J. R., Feldman, U., Tsuneta, S., & Uchida, Y.  1992, PASJ, 44, L181

Kobayashi, K., Cirtain, J., Winebarger, A. R., et al.  2014, SoPh, 289, 4393

Kucera, T. A., Young, P. R., Klimchuk, J. A., & DeForest, C. E.  2019, ApJ, 885, 7

Lopez Fuentes, M. C., Demoulin, P., & Klimchuk, J. A.  2008, ApJ, 673, 586

Lopez Fuentes, M. C., Klimchuk, J. A., & Demoulin, P.  2006, ApJ, 639, 459

Malanushenko, A., & Schrijver, C. J.  2013, ApJ, 775, 120

Malanushenko, A., Rempel, M., Cheung, M., DeForest, C., & Klimchuk, J. A. 2020, in preparation.

Patsourakos, S., Klimchuk, J. A., & MacNeice, P. J.  2004, ApJ, 603, 322

Peter, H., & Bingert, S.  2012, AA, 548, A1

Peter, H., Bingert, S., Klimchuk, J. A., et al.  2013, AA, 556, A104

Porter, L. J., & Klimchuk, J. A.  1995, ApJ, 454, 499

Rachmeler, L. A.  2019, SoPh, 294, 174

Rempel, M.  2017, ApJ, 834, 10

Shelyag, S., Keys, P., Mathioudakis, M., & Keenan, F. P.  2011, AA, 526, A5

Thalmann, J. K., Tiwari, S. K., & Wiegelmann, T.  2014, ApJ, 780, 102

Toriumi, S., & Wang, H.  2019, Liv Rev Sol Phys, 16, 3

Van Driel-Gesztely, L., & Green, L. M.  2015, Liv. Rev. Sol. Phys., 12, 1

Watko, J. A., & Klimchuk, J. A.  2000, SoPh, 193, 77

Welsch, B. T.  2015, PASJ, 67 (2), 18

West, M., Zhukov, A., & Klimchuk, J.  2014, in 40[th] COSPAR Scientific Assembly 40

Williams, T., Walsh, R. W., Winebarger, A. R., et al.  2020, ApJ, 892, 134

Winebarger, A. R., Cirtain, J., Golub, L., et al.  2014, ApJL, 787, L10

Yates, A. R., Hornig, G., & Welsch, B. T.  2012, AA, 539, A1